\documentclass[11pt]{article}
\usepackage{amssymb,amsmath,amsfonts}
\usepackage{graphicx}
\usepackage{graphics}

\begin{document}

\title{Solution of the Klein-Gordon equation in external Yang-Mills gauge
field}
\author{V. V. Parazian\thanks{%
E-mail: vparazian@gmail.com } \\
\textit{Institute of Applied Problems of Physics NAS RA,}\\
\textit{25 Nersessian Street, 0014 Yerevan, Armenia}}
\maketitle

\begin{abstract}
Exact solutions of the Klein-Gordon equation in an external non-Abelian
gauge field with an $SU(N)$ symmetry group have been obtained. The external
field is a solution of the Yang-Mills equations and describes a plane wave
on the light cone. The obtained solutions form a complete set and can be
used in the procedure of canonical quantization of scalar fields.
\end{abstract}

\bigskip

\section{Introduction}

The modern theory of elementary particles (the standard model) is formulated
within the framework of quantum field theory. An important component of this
model is the Yang-Mills gauge fields, which describe the weak and strong
interactions between elementary particles. Initially introduced into physics
for this purpose, models with Yang-Mills fields have now found applications
in condensed matter physics, as well as in atomic and optical systems (see,
for example, \cite{Berche2013,Tan2020} and references therein). Interesting
examples of applications are spin superfluids, Mott insulators, quantum spin
transport in nanoscale systems of semiconductor nanostructures with strong
spin-orbit coupling.

Yang-Mills fields are nonlinear, and therefore the corresponding quantum
theory is quite complicated. In a number of problems both in particle
physics and in effective models describing the properties of a number of
systems in condensed matter physics, external fields can be considered
classically with fairly good accuracy (a review of classical solutions in
quantum field theory can be found in \cite{Weinberg,Weinberg2012}). This
approach is widely used in quantum electrodynamics and in quantum field
theory in curved space-time to model the effect of the gravitational field
on quantum effects. In particular, classical solutions of the equations of
motion of Yang-Mills fields play an important role in the analysis of
nonperturbative phenomena, confinement, topological properties, and
questions of vacuum stability. Well-known examples of classical solutions in
chromodynamics are monopoles, instantons, and solitons.

Simpler are scalar fields that describe particles with zero spin. The
best-known fundamental field of this kind is the Higgs field in the standard
model of elementary particles. Scalar fields play an important role in
cosmology. In particular, in most inflationary models of the expansion of
the early Universe, the accelerating expansion is driven by a scalar field
called the inflaton. Scalar fields also appear in the effective description
of phonon degrees of freedom in condensed matter physics.

In the presented work, solutions of the equation of motion of a scalar field
in an external Yang-Mills field are investigated. As an external field, the
solution of the classical Yang-Mills equations obtained in \cite{Coleman1977}
is chosen (for wave solutions of the Yang-Mills equations, see, for example,
\cite{Nutku1983}-\cite{Rabinowitch2023} and the references therein). This
solution is a non-Abelian analog of plane electromagnetic waves. The
corresponding solution of the Dirac equation is considered in \cite%
{Koshelkin}. Compared with the fermion case, the analysis of systems with
interacting scalar and Yang-Mills fields is simpler, which makes these
systems an ideal laboratory for analyzing the confinement mechanism in
numerical studies of functional equations, as well as in lattice models \cite%
{Fister2010}. Classical solutions of the Yang-Mills field equations in the
presence of massless scalar fields with fourth-degree self-interaction are
considered in \cite{Vinet1981}. When introducing scalar matter fields into a
theory with Yang-Mills fields, a fourth-degree interaction term between the
scalar fields and the Faddeev-Popov ghosts naturally arises \cite{Capri}.

The paper is organized as follows. In the next section, the Klein-Gordon
equation in an external Yang-Mills gauge field is considered. The structure
of the external field is described. The solution of the scalar field
equation in an external non-Abelian gauge field is presented in Section 3.
The main results of the paper are summarized in Section 4.

\section{Klein-Gordon equation with Yang-Mills gauge field}

Consider a scalar field $\phi \left( x\right) $ in an external non-Abelian
gauge field $A_{a}^{\mu }\left( x\right) $ corresponding to a group $SU(N)$.
The boson field $\phi \left( x\right) $ and the Yang-Mills gauge field $%
A_{a}^{\mu }\left( x\right) $ respectively, constitute the space of the
fundamental ($\phi \left( x\right) $) and associated ($A_{a}^{\mu }\left(
x\right) $) -representations of the group $SU(N)$. The equation of motion
for the field $\phi \left( x\right) $ reads

\begin{equation}
\left( g^{\mu \nu }D_{\mu }D_{\nu }+m^{2}\right) \phi =0,  \label{KGeq}
\end{equation}%
where $m$ is the mass of the field quanta, $g^{\mu \nu }=\mathrm{diag}%
(1,-1,-1,-1)$ is the metric tensor, $D_{\mu }=\partial _{\mu }+igA_{\mu
}^{a}\left( x\right) T_{a}$ is the gauge extended covariant derivative, $%
T_{a}$, $a=1,\ldots N^{2}-1$, are the generators of the $SU(N)$ group, and $%
g $ is the coupling constant. The gauge field obeys the equation
\begin{equation}
\partial _{\mu }F_{a}^{\nu \mu }\left( x\right) -g\,f_{ab}\,^{c}\,A_{\mu
}^{b}\left( x\right) F_{c}\,^{\nu \mu }\left( x\right) =0,  \label{YMeq}
\end{equation}%
with the field tensor%
\begin{equation}
F_{a}^{\nu \mu }\left( x\right) =\partial ^{\nu }A_{a}^{\mu }\left( x\right)
-\partial ^{\mu }A_{a}^{\nu }\left( x\right) -g\,f_{a}\,^{bc}A_{b}^{\nu
}\left( x\right) A_{c}^{\mu }\left( x\right) ,  \label{Fnyumyudefin}
\end{equation}%
and $f_{ab}\,^{c}$ being the structure constants of the $SU(N)$ group. In
the formulae above summation over any pair of repeated indexes is assumed.

The generators $T_{a}$ obey the commutation relation%
\begin{equation}
\left[ T_{a},T_{b}\right] _{-}=T_{a}T_{b}-T_{b}T_{a}=if_{ab}\,^{c}T_{c},
\label{commut}
\end{equation}%
and the normalization condition
\begin{equation}
\mathrm{Tr}\left( T_{a}T_{b}\right) =\frac{1}{2}\delta _{ab}.  \label{TrTT}
\end{equation}%
The structure constants $f_{ab}\,^{c}$ are real and anti-symmetric with
respect to all indexes. From (\ref{commut}) and (\ref{TrTT}) we get the
relation for the anticommutator $T_{a}T_{b}+T_{b}T_{a}\equiv \left[
T_{a},T_{b}\right] _{+}$:

\begin{equation}
\left[ T_{a},T_{b}\right] _{+}=\frac{\delta _{ab}}{N}+d_{abc}T^{c}.
\label{anticommut}
\end{equation}%
Here, the coefficients $d_{abc}$, defined by
\begin{equation}
d_{abc}=\mathrm{Tr}\left( \left[ T_{a},T_{b}\right] _{+}T_{c}\right) ,
\label{dabc}
\end{equation}%
are real and symmetric with respect to any pair of indices.

We will consider a special gauge field configuration corresponding to a
plane wave with the 4-wavevector $k=(k^{0},\mathbf{k})$ \cite{Coleman1977},
obeying the condition $k^{\mu }k_{\mu }=0$. Introducing the notation $%
\varphi =k\,x$, the vector potential for that configuration has the form
\begin{equation}
A_{\mu }^{a}\left( x\right) =A_{\mu }^{a}\left( \varphi \right) .
\label{Aplanewave}
\end{equation}%
Assuming that the wave propagates along the negative direction of the $x^{3}$%
-axis, it is convenient to introduce light cone coordinates $x^{\pm
}=x^{0}\pm x^{3}$. In these coordinates $A_{-}^{a}=A_{1}^{a}=A_{2}^{a}=0$
and the gauge field has a single nonzero component $A_{+}^{a}\left( x\right)
$. The latter is presented in the form (see \cite{Coleman1977})
\begin{equation}
A_{+}^{a}\left( x\right) =f^{a}\left( x^{+}\right) x^{1}+g^{a}\left(
x^{+}\right) x^{2}+h^{a}\left( x^{+}\right) ,  \label{A Coleman solut}
\end{equation}%
with $f^{a}\left( x^{+}\right) $, $g^{a}\left( x^{+}\right) $, and $%
h^{a}\left( x^{+}\right) $ being arbitrary bounded functions. By a suitable
gauge transformation, we can impose the axial gauge condition
\begin{equation}
\partial ^{\mu }A_{\mu }^{a}\left( x\right) =k^{\mu }\dot{A}_{\mu
}^{a}\left( x\right) =\partial ^{+}A_{+}^{a}\left( x\right) =0,
\label{axial gauge}
\end{equation}%
from which it follows that $k^{\mu }A_{\mu }^{a}\left( x\right) =0$. Here
and below the dot stands for the derivative with respect to $\varphi $. It
can be checked that the gauge field configuration described above obeys the
field equation (\ref{YMeq}).

\section{Solution in a plane wave background}

By using the relations for the non-Abelian gauge field, given above, the
Klein-Gordon equation (\ref{KGeq}) is written in the form
\begin{equation}
\left[ \partial ^{\nu }\partial _{\nu }+2igA^{\nu a}\left( x\right)
T_{a}\partial _{\nu }-\frac{g^{2}}{2N}A^{\nu a}\left( x\right) A_{\nu
a}\left( x\right) +m^{2}\right] \phi \left( x\right) =0.
\label{simplifyKGeq}
\end{equation}%
We are interested in the solutions of this equation having the form $\phi
\left( x\right) =\phi _{\alpha }\left( x,p\right) $, where the function $%
\phi _{\alpha }\left( x,p\right) $ has the structure
\begin{equation}
\phi _{\alpha }\left( x,p\right) =e^{-ipx}F_{\alpha }\left( k\,x\right) ,
\label{generalformsolut}
\end{equation}%
with a 4-momentum $p^{\nu }=\left( p^{0},\mathbf{p}\right) $. Here, $%
F_{\alpha }\left( k\,x\right) $ is a multicomponent function that depends on
the variable $\alpha $ taking integer values from $1$ to $N$. The latter
specifies the state of a scalar field in the space of the fundamental
representation of the $SU(N)$ group.

Substituting the $\phi _{\alpha }\left( x,p\right) $ from (\ref%
{generalformsolut}) into the field equation (\ref{simplifyKGeq}) and using (%
\ref{axial gauge}), the following equation is obtained for the function $%
F_{\alpha }\left( k\,x\right) $:%
\begin{equation}
2ipk\dot{F}_{\alpha }\left( k\,x\right) =\left[ -\left( p^{2}-m^{2}\right) -%
\frac{g^{2}}{2N}A_{\nu }^{a}\left( x\right) A_{a}^{\nu }\left( x\right)
+2gp^{\nu }A_{\nu }^{a}\left( x\right) T_{a}\right] F_{\alpha }\left(
k\,x\right) ,  \label{KGeqforF}
\end{equation}%
where $pk=p^{\mu }k_{\mu }$. The formal solution of this equation reads%
\begin{equation}
F_{\alpha }\left( kx\right) =\exp \left\{ \frac{i}{2pk}\left[ \left(
p^{2}-m^{2}\right) \varphi +\frac{g^{2}C(\varphi )}{2N}-2gB^{a}(\varphi
)T_{a}\right] \right\} v_{\alpha }.  \label{formalsolut}
\end{equation}%
where $v_{\alpha }$ is a constant spinor that is an element of the space of
the corresponding representation and we have used the notations%
\begin{eqnarray}
B^{a}\left( \varphi \right) &=&\int_{0}^{\varphi }d\varphi ^{\prime }p^{\nu
}A_{\nu }^{a}\left( \varphi ^{\prime }\right) ,  \notag \\
C(\varphi ) &=&\int_{0}^{\varphi }d\varphi ^{\prime }A_{\nu }^{a}\left(
\varphi ^{\prime }\right) A_{a}^{\nu }\left( \varphi ^{\prime }\right) .
\label{Ba}
\end{eqnarray}

The part in (\ref{formalsolut}) with the last term in the square brackets is
further simplified expanding the exponent and using the commutation
relations for the matrices $T_{a}$. The results is presented as
\begin{equation}
\exp \left[ -i\frac{g}{pk}B^{a}\left( \varphi \right) T_{a}\right] =\left[ 1-%
\frac{ig}{pk}\frac{\tan \theta }{\theta }B^{a}\left( \varphi \right) T_{a}%
\right] \cos \theta ,  \label{expseries1}
\end{equation}%
where we have defined%
\begin{equation}
\theta =\frac{g}{pk}\left[ \frac{B^{a}\left( \varphi \right) B_{a}\left(
\varphi \right) }{2N}\right] ^{\frac{1}{2}}.  \label{teta}
\end{equation}%
By taking into account the relation (\ref{expseries1}), the solution (\ref%
{formalsolut}) is expressed as
\begin{eqnarray}
\phi _{\alpha }\left( x,p\right) &=&e^{-ipx}\exp \left\{ \frac{i}{2pk}\left[
\left( p^{2}-m^{2}\right) \varphi +\frac{g^{2}C(\varphi )}{2N}\right]
\right\}  \notag \\
&&\times \cos \left( \theta \right) \left[ 1-\frac{ig}{pk}\frac{\tan \theta
}{\theta }B^{a}\left( \varphi \right) T_{a}\right] v_{a}.  \label{KGsolution}
\end{eqnarray}%
The spinor $v_{\alpha }$ is normalized by the condition%
\begin{equation}
v_{\alpha }^{\dag }v_{\beta }=\delta _{\alpha \beta }.
\label{spinor v defin}
\end{equation}

The function $\phi _{\alpha }\left( x,p\right) $ is an exact solution of the
Klein-Gordon equation in the external non-Abelian field, which satisfies the
Yang-Mills equations and can be written in the form of a plane wave on the
light cone. The function (\ref{KGsolution}) is normalized by the condition%
\begin{equation}
\int d^{3}x\phi _{\alpha }^{\ast }\left( x,p^{\prime }\right) \phi _{\alpha
}\left( x,p\right) =\left( 2\pi \right) ^{3}\delta \left( p-p^{\prime
}\right) .  \label{normalcond}
\end{equation}

It can be checked that the functions $\phi _{\alpha }\left( x,p\right) $ and
$\phi _{\alpha }\left( x,-p\right) $ are orthogonal. The solutions $\phi
_{\alpha }\left( x,p\right) $ present positive frequency modes, while $\phi
_{\alpha }\left( x,-p\right) $ are negative frequency modes \cite{Berest1981}%
, \cite{Weinberg}. With these solutions, the general solution of the
Klein-Gordon equation, describing both particles and antiparticles, can be
written in the form of the expansions%
\begin{eqnarray}
\phi \left( x\right) &=&\sum_{\alpha }\int \frac{d^{3}p}{\sqrt{2p^{0}}\left(
2\pi \right) ^{3}}\left[ \hat{a}_{\alpha }\left( p\right) \phi _{\alpha
}\left( x,p\right) +\hat{b}_{\alpha }^{\dag }\left( p\right) \phi _{\alpha
}\left( x,-p\right) \right] ,  \notag \\
\phi ^{\dag }\left( x\right) &=&\sum_{\alpha }\int \frac{d^{3}p}{\sqrt{2p^{0}%
}\left( 2\pi \right) ^{3}}\left[ \hat{a}_{\alpha }^{\dag }\left( p\right)
\phi _{\alpha }^{\dag }\left( x,p\right) +\hat{b}_{\alpha }\left( p\right)
\phi _{\alpha }^{\dag }\left( x,-p\right) \right] ,  \label{generalsolutKG}
\end{eqnarray}%
Here, the symbols $\hat{a}_{\alpha }^{\dag }\left( p\right) $, $\hat{b}%
_{\alpha }^{\dag }\left( p\right) $ and $\hat{a}_{\alpha }\left( p\right) $,
$\hat{b}_{\alpha }\left( p\right) $ denote the creation and annihilation
operators of scalar particles $\left( \hat{a}_{\alpha }\left( p\right) ,\hat{%
a}_{\alpha }^{\dag }\left( p\right) \right) $ and antiparticles $\left( \hat{%
b}_{\alpha }\left( p\right) ,\hat{b}_{\alpha }^{\dag }\left( p\right)
\right) $. These operators satisfy the standard commutation relations for
bosonic fields.

\section{Conclusion}

We have discussed the exact solutions of the Klein-Gordon scalar field
equation in an external Yang-Mills gauge field describing a plane wave on a
light cone. In light cone coordinates, the only nonzero component of the
gauge field is given by expression (\ref{A Coleman solut}). The solutions
found describe plane waves of a scalar field. They form a complete system of
solutions of the Klein-Gordon equation and can therefore be used in the
procedure of canonical quantization of a scalar field in an external
classical Yang-Mills field. They can also be used to evaluate the Green's
function of a scalar field, which plays an important role in quantum field
calculations of various processes in a system of interacting scalar and
Yang-Mills fields.

\end{document}